# Large Language Models (GPT) for automating feedback on programming assignments

**Maciej PANKIEWICZ[a]\* & Ryan S. BAKER[b]**
[a]*Institute of Information Technology, Warsaw University of Life Sciences, Poland*
[b]*Penn Center for Learning Analytics, University of Pennsylvania, USA*
\*maciej_pankiewicz@sggw.edu.pl

**Abstract:** Addressing the challenge of generating personalized feedback for programming assignments is demanding due to several factors, like the complexity of code syntax or different ways to correctly solve a task. In this experimental study, we automated the process of feedback generation by employing OpenAI's GPT-3.5 model to generate personalized hints for students solving programming assignments on an automated assessment platform. Students rated the usefulness of GPT-generated hints positively. The experimental group (with GPT hints enabled) relied less on the platform's regular feedback but performed better in terms of percentage of successful submissions across consecutive attempts for tasks, where GPT hints were enabled. For tasks where the GPT feedback was made unavailable, the experimental group needed significantly less time to solve assignments. Furthermore, when GPT hints were unavailable, students in the experimental condition were initially less likely to solve the assignment correctly. This suggests potential over-reliance on GPT-generated feedback. However, students in the experimental condition were able to correct reasonably rapidly, reaching the same percentage correct after seven submission attempts. The availability of GPT hints did not significantly impact students' affective state.

**Keywords:** Programming, automated assessment tools, automated feedback, LLM, GPT

## 1. Introduction

Large Language Models (LLMs) are deep neural network models able to effectively process and analyze complex linguistic structures (Carlini et al., 2021). LLMs have garnered significant attention in the field of natural language processing (NLP) because of the ability to generate human-like text. This feature makes LLMs a promising technology in educational settings (Finnie-Ansley et al., 2022; Dai et al., 2023; Pardos & Bhandari, 2023), where the provision of personalized feedback is integral to scaffold learning effectively (Jackson & Graesser, 2007; Hull & du Boulay, 2015).

One of the most prominent examples of LLMs is the OpenAI's Generative Pre-trained Transformer (GPT) series (Radford et al. 2019; Brown et al. 2020), which leverages a transformer architecture and is able to capture long-range dependencies and positional information within a text. It is available through the ChatGPT application and via an Application Programming Interface (API), enabling integration with different existing applications, including educational apps. Especially in the field of computer science education, positive effects of such embedment may emerge earlier than in other educational domains, because of the significant effort conducted to develop models trained on large sets of programming code examples (Finnie-Ansley et al., 2022).

The field of computer science education has already extensively leveraged software tools that automate the process of self-paced learning. Multiple automated assessment tools have been designed to evaluate and provide feedback on student performance in various educational tasks (Deeva et al. 2021). The proliferation of these tools in computer science education, in areas from introductory programming (Edwards & Murali, 2017; Brusilovsky et al., 2018) to databases (Stanger, 2018), has had benefits and has contributed to the development of self-paced learning environments in other domains (Paiva et al., 2022). Along with this development, there is a growing body of research on personalizing feedback to enhance student learning outcomes (Deeva et al. 2021). Nevertheless, effective implementation of this approach continues to present significant challenges, given the difficulty of comprehensively assessing student work (Maier & Klotz, 2022).

In this study, we try to leverage the potential of GPT models for supporting students in providing feedback on programming tasks by integrating GPT through an API with an application for automated assessment of programming code. The GPT-3.5 model is used to automatically generate personalized feedback for university students taking an *Object-oriented programming* course. In this paper, we presented a controlled experiment comparing a system that adds GPT-generated hints to a system that only uses the original, human-developed hints. We compare the two conditions across multiple dimensions, including immediate performance, usage, performance on later content (without GPT-generated hints), time taken to complete assignments, and affect.

## 2. Methods

The study involved second-semester computer science students at the Warsaw University of Life Sciences (Poland) enrolled in an *Object-oriented programming* course that was required for their major. The course was conducted using the `C#` programming language. A total of 132 students consented to participate (out of 174 students taking the course) and were randomly assigned to either the control (N=66) or the experimental group (N=66). A pre-test was administered at the beginning of the semester to establish a baseline understanding of the students' knowledge before engaging with the platform's content. The process of data collection spans the initial part of the semester, specifically covering a nine-week period within the fifteen-week academic term. This timeframe extends from the first week of March 2023 through to the first week of May 2023. 93 students submitted at least one solution during the period of the study (control: 46, experimental: 47).

The experiment utilized the *RunCode* online application – a platform for automated execution and testing of a programming code available at runcodeapp.com (Pankiewicz, 2020). The application has been used by students at the University since 2017 within programming courses. All students that participated in this experiment were familiar with the application and had actively used it during the *Introduction to programming* course taught in the previous (first) semester. Students submitted programming code using an editor integrated within the application.

46 programming assignments covering basic object-oriented programming concepts (classes, objects, fields, methods, constructors, encapsulation, inheritance, and polymorphism) have been made available on the platform for the purpose of practicing the course material. These concepts were introduced during the first 6 weeks of the course. Each assignment was composed of a collection of more specific subtasks, such as: the creation of a class, the addition of fields with appropriate access modifiers, the definition of constructors, etc. A comprehensive suite of 809 unit tests was developed to thoroughly assess performance on these subtasks, ensuring a detailed evaluation of their individual components. Multiple submissions on a task were allowed. Usage of the platform was voluntary, and neither usage nor results within the platform counted towards the final grade. 5923 code submissions were collected during the period of study (control: 3077, experimental: 2846).

The unit tests were designed to validate code requirements specified by the assignment. For instance, a test aimed at confirming the presence of the "User" class within the submitted code might be denoted as `TestIsClassDefined("User")`, with an expected value of `true`. This approach ensured that different aspects of the assignment's requirements were thoroughly examined and verified in the students' code submissions.

The application provided students with a score (0-100%) and feedback after each code submission. If the submitted code failed to compile, the regular feedback available on the platform was presented without requiring any additional clicks and included details on compiler errors. The platform also highlighted in the online code editor lines where errors occurred. If the code compiled successfully, the feedback contained information on the executed unit tests, including input values and expected output. Students needed to click on a specific test from the list to access detailed information about its execution. The tests were color-coded, with green indicating success and red signifying failure, to facilitate easy identification of the test outcomes.

Both the control and experimental groups had access to the same set of tasks within the application. However, for 38 out of 46 tasks, students in the experimental group had an additional feature enabled, which provided them with feedback generated by a large language model through the GPT-3.5 API. For the remaining 8 tasks, students in the experimental group only had access to the regular feedback offered by the platform. These tasks were slightly more challenging and were designed to encapsulate the concepts introduced in previously solved tasks, as well as assess the comprehension of these concepts.

Within the experimental condition, additional feedback was provided by the GPT model, going beyond the standard feedback provided by the application, which included information on compiler errors, runtime errors and details on executed unit tests. The GPT-generated feedback provided suggestions for code improvement, explanations of compiler errors, and hints for debugging. The feedback was designed to be informative and constructive, aiming to guide students towards correct solutions and improve their understanding of programming concepts, without revealing the correct code solution. This feedback was presented immediately after code submission, with a prompt asking students to rate the usefulness of the hint on a 5-point Likert scale ranging from 'Not useful at all' to 'Extremely useful'. Feedback was automatically generated in Polish, to match the assignments and the interface of the system, which also used this language.

Feedback was requested from GPT when the code submitted by a student exhibited compiler errors, runtime errors, or failed at least one unit test. The dynamically generated prompt included the assignment text (in Polish) and the student's code, accompanied by additional information based on the testing results. This information contained a compiler message for non-compiling code, exception type and message for compiled code that generated an exception during execution, or details regarding a failed unit test: the test name, input values, and expected outcome. To accommodate the OpenAI GPT API's token limitation (4,000 tokens), assignments were designed in a way that ensured that even the most complex tasks and accompanying code could fit within the imposed constraint. The prompt text was similar for the three scenarios mentioned, with variations stemming from the testing process outcomes. In cases where the code did not compile, a specific prompt was generated:

```
I want you to act as a Stackoverflow post that helps me to solve a programming
assignment in C#. I want you to explain in Polish why this code does not compile.
Don't write solution in the explanation, but focus on meaningful hints. I want you
to also include the line where the compiler error occurred in the explanation. I
want you to also include a line number for each detected error in the explanation.
To help me better understand your response, highlight keywords, line numbers, class
names, variable names, messages, line numbers and error names with the <code> </code>
HTML markup in the explanation. Programming assignment: ### <ASSIGNMENT_TEXT> ###
C# code: <STUDENT_CODE> ### Compiler errors: <COMPILER_ERRORS>
```

The following parameters were used for the request: 'model': 'text-davinci-003', 'temperature': 0, 'max_tokens': 500, 'n': 1, 'top_p': 1, 'frequency_penalty': 0, 'presence_penalty': 0.

An illustration of the hint proposed by the GPT in the event of encountering code with a syntax error is presented as follows (translated from Polish to English with highlights generated by the GPT): "A compiler error `error CS1002` occurred in line 5 of the code with the message `; expected`. This means that there is a missing semicolon `;` after the declaration of the `string name` field in line 5. To fix the error, a semicolon `;` should be added at the end of line 5. A `;` is required at the end of every declaration in C#."

Students self-reported their affective states while completing tasks on the platform by interacting with a dynamic HTML element. The platform prompted students for their emotional state following the receipt of submission results by asking them to select the option that best described their current feelings. The response options included: *Focused*, *Anxious*, *Bored*, *Confused*, *Frustrated*, and *Other* (in this order), accompanied by representative emoticons. These affective states were chosen due to their relevance to the learning process and their prevalence in past research (Karumbaiah et al., 2022). To mitigate potential frustration from overly frequent survey prompts, the platform randomly decided after each submission whether to present the survey, with a probability of one in three.

## 3. Results

Due to violations of normality assumptions, non-parametric Mann-Whitney U tests were utilized. Benjamini-Hochberg alpha correction (Benjamini & Hochberg, 1995) was applied to control for multiple comparisons in analysis of differences in affective state frequencies among students.

### 3.1 Pre-test results

To ensure that the control and experimental groups were comparable in terms of their initial understanding of the relevant programming concepts at the beginning of the study, a pretest consisting of 7 multiple-choice questions was administered to both the control and experimental groups. These questions covered key topics on object-oriented programming, including classes, constructors, encapsulation, inheritance, and polymorphism. No statistically significant differences were identified between the control (Mdn=0) and the experimental (Mdn=1) groups (W=1791, p=0.249, Mann-Whitney U test).

### 3.2 Usefulness of the GPT hints

In the examination of the perceived usefulness of the GPT generated hints, a total of 1,442 responses were collected from N=46 participants within the experimental group throughout the duration of the study. As the control group did not have access to the GPT hints, no responses were obtained from this group. Figure 1 displays the histogram of the frequency of users according to the median rating they gave to received hints, with the x-axis representing the median of the hint rating for a user.

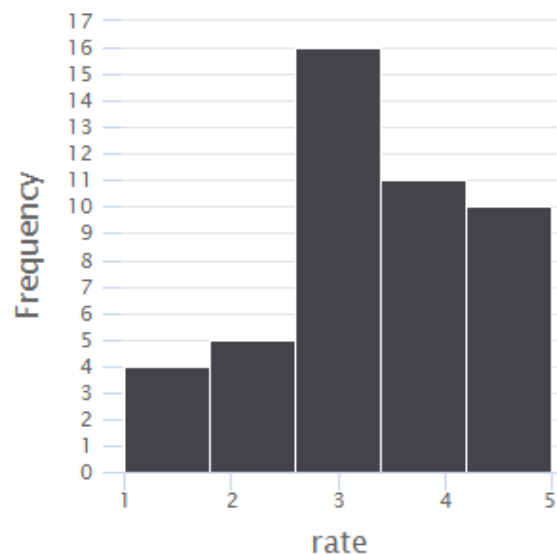

*Figure 1.* Distribution of users according to the median rating for received hints.

As Figure 1 shows, the majority of users rated the GPT hints they received as useful – these results suggest that the hint feature was generally well-received by users.

### 3.3 Usage of the regular platform feedback

One possible impact of the GPT feedback is that learners may use the other feedback less often. To evaluate whether there were differences in the usage of the regular (non-GPT) platform feedback between the experimental and control groups, we analyzed the percentage of incorrect submissions for which students clicked on feedback to access details about tests that failed. We did not examine clicks on feedback for tests that ended successfully, as the application did not generate GPT hints for these tests. Therefore, students who only made successful submissions, and students who submitted only non-compiling code in their submissions (in this case the platform always presents feedback, so users are not required to click on it), were not included in this analysis.

A statistically significant difference was found between the experimental (Mdn=0.321, N=45) and control (Mdn=0.710, N=45) groups (W=282, p<0.001, Mann-Whitney U test) for the set of tasks where experimental group had GPT feedback enabled (38 out of 46). The experimental group used the regular feedback significantly less than the control group for these tasks.

However, no significant difference was found between the experimental (Mdn=0.769, N=23) and control (Mdn=0.667, N=25) groups (W=341, p=0.269, Mann-Whitney U test) for the set of tasks, where GPT feedback was not available for experimental group (8 out of 46). When GPT feedback was unavailable, the experimental group utilized the platform's regular feedback in a similar amount to the control group (Figure 2).

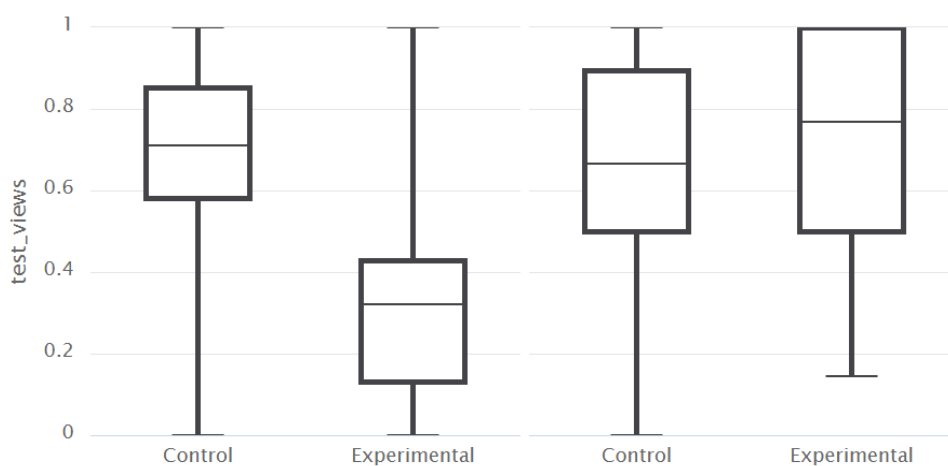

*Figure 2.* Comparison of the percentage of incorrect submissions with at least one click to request (view) details on the test that failed (regular platform's feedback) between the control and experimental group (on the left: tasks with GPT feedback enabled for experimental group; on the right: tasks with GPT feedback disabled for the experimental group).

### 3.4 Performance within the platform – GPT hints enabled

In order to examine the influence of the GPT hints on student performance within the platform, we conducted an analysis of users' consecutive task attempts for tasks where GPT hints were enabled for the experimental group (38 out of 46 tasks). The primary focus was on assessing the percentage of successful submissions after each attempt (cumulative).

For the first attempt, the experimental group appeared to have a similar percentage of successes (48.3%) as the control group (48.0%). Values for the second and all consecutive attempts decreased for both groups, with performance worsening across attempts. Following the second attempt, the experimental group appeared to have a slightly higher percentage of successful submissions, a trend that persisted across subsequent attempts. For the purpose of clarity, only the first 15 attempts are depicted in the chart (Figure 3).

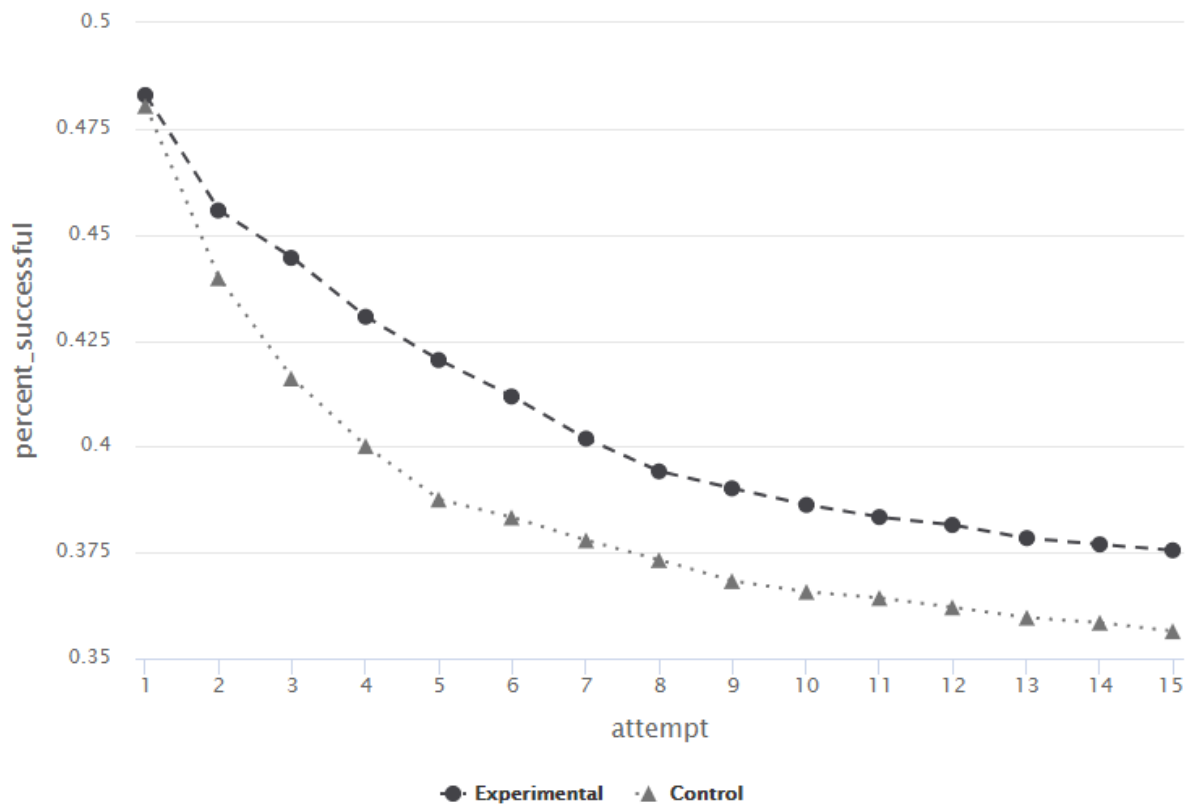

*Figure 3.* Comparison of the percentage of successful submissions at each attempt (cumulative) for the control and experimental group, for tasks where GPT hints were enabled in the experimental group.

We employed a linear mixed-effects model to test whether there was a significant difference in slopes between the control and experimental groups concerning the number of attempts. The model was fit using restricted maximum likelihood (REML) estimation with the `nlme` R package (Pinheiro et al., 2017). The fixed effects of the model consisted of the main effects of group and attempt, as well as the interaction between group and attempt. Our analysis revealed a non-significant interaction effect between group and attempt, $t(13)=-0.172$, $p=0.866$. This indicates that the slopes of the control and experimental groups do not significantly differ. Nonetheless, the main effects of group and attempt were found to be statistically significant, group: $t(13)=5.38$, $p<0.001$; attempt: $t(13)=-7.38$, $p<0.001$. This finding suggests that, averaged across all attempts, the experimental group scored higher than the control group on the dependent variable.

*3.5 Performance within the platform – without GPT hints*

To further evaluate the impact of the introduced platform feature on the experimental group, we analyzed performance of both groups on the tasks where GPT-generated hints were not provided for the experimental group (8 out of the total 46 tasks).
As in the prior section, we visualized the percentage of successful submissions following each attempt. For the initial six attempts, the control group performed better than the experimental group; however, starting from the seventh attempt correctness rates were comparable (Figure 4).
A linear mixed-effects model fit using restricted maximum likelihood (REML) was employed to assess whether the slopes of the control and experimental groups were significantly different with respect to the number of attempts. The model revealed a significant interaction effect between group and attempt, t(13)=7.13, p<0.001. This finding indicates that the slopes of the control and experimental groups were significantly different.

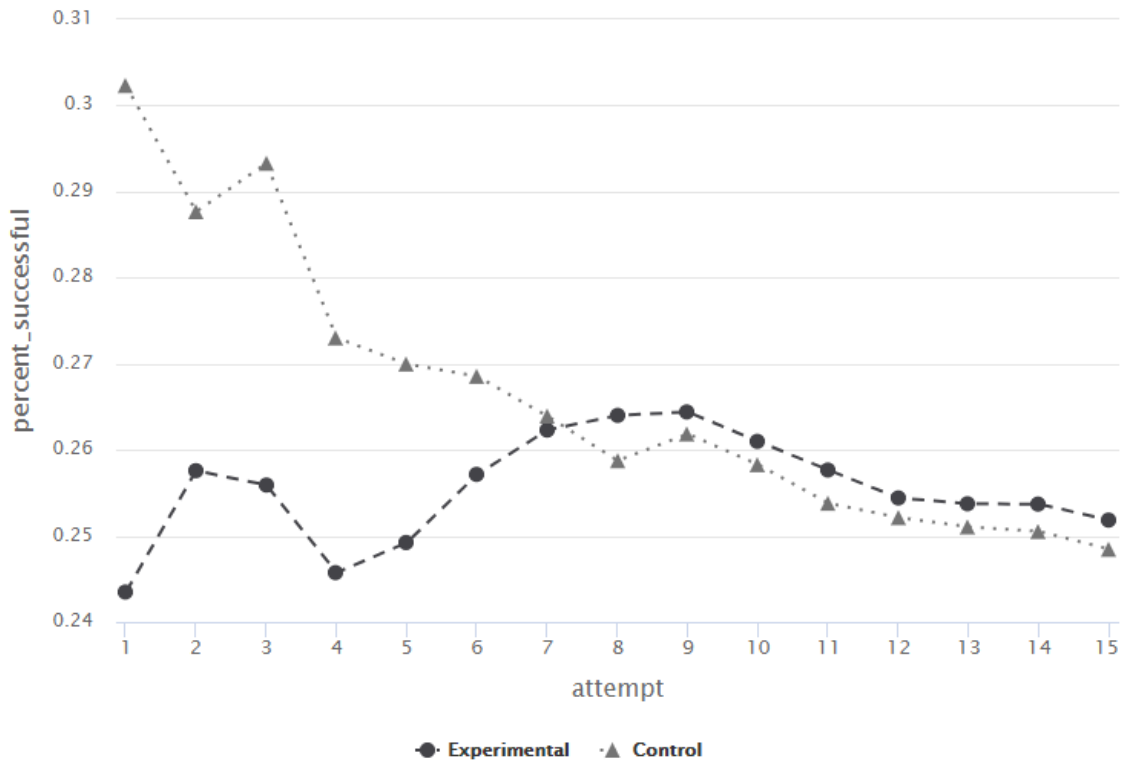

*Figure 4.* Comparison of the percentage of successful submissions at each attempt (cumulative) for the control and experimental group for tasks where GPT hints were disabled in the experimental group.

*3.6 Time needed to successfully complete tasks*

In order to assess the differences in the time required for students in the experimental and control groups to successfully complete tasks on the platform, we analyzed 1,784 successful student submissions during the study period (control: 905, experimental: 879).

We evaluated the time taken by students in both conditions to fully solve each task, while excluding tasks completed by fewer than three students in each condition. Due to the platform's lack of keystroke-level data monitoring, a time limit of 7,200 seconds (2 hours) was established for tasks not solved within this duration. This decision was made because some students ceased working on tasks without logging out.

A significant majority (88%) of successful student attempts were completed within this time frame. Due to violations of normality assumptions, we used rank-based regression – a non-parametric alternative to traditional likelihood or least squares estimators (Kloke & McKean, 2012).

We analyzed the time performance on tasks where the GPT feedback was available for the experimental group. For this set of tasks, the group was not a statistically significant predictor of the time needed to solve tasks, t(89)=-0.24, p=0.811. However, the pretest score was a marginally significant predictor, t(89)=-0.69, p=0.095.

We also conducted the analysis for tasks where the GPT feedback was not available for the experimental group. The group was a statistically significant predictor of the time needed to solve tasks, with the experimental group needing less time than the control group, t(48)=-2.25, p=0.029. The pretest score was not significantly associated with time needed to solve tasks, t(48)=-0.36, p=0.723. For the set of these tasks results indicate that students in the experimental group took, on average, 375 seconds (6.25 min.) less time than the control group, controlling for the pretest score.

*3.7 Affective states*

A total of 1,304 affect survey responses were collected from participants over the course of the study (645 in the experimental condition, 659 control). In the analysis, we include students who submitted at least 3 responses in this survey: in the control N=33, and in the experimental group: N=35.

Both conditions predominantly reported the *focused* state, which constituted over half of the responses in each group. No significant differences between conditions were observed for the reported frequency of any of the affective states (Table 1).

Table 1. *Affective State Survey – Statistical Evaluation of Differences in Frequencies of Affective States Reported by Students in Experimental and Control Group*

|  | Experimental (Mdn/Mean) | Control (Mdn/Mean) | W statistic | p-value |
|---|---|---|---|---|
| Focused | 0.538/0.52 | 0.5/0.54 | 537 | 0.620 |
| Frustrated | 0/0.14 | 0/0.13 | 611 | 0.653 |
| Anxious | 0/0.12 | 0/0.12 | 550.5 | 0.718 |
| Confused | 0/0.06 | 0/0.09 | 571 | 0.935 |
| Bored | 0/0.03 | 0/0.05 | 559.5 | 0.782 |

*Frustration* did not significantly differ between the control (Mdn=0) and experimental (Mdn=0) groups (W=611, adjusted α=0.02, p=0.653, Mann-Whitney U test).

Similarly, no significant difference was found for *boredom* between the experimental (Mdn=0) and control (Mdn=0) groups (W=559.5, adjusted α=0.04, p=0.782, Mann-Whitney U test).

The same pattern was observed for *anxiety* and *confusion*, with no significant difference between the experimental (Mdn=0) and control (Mdn=0) groups for anxiety (W=550.5, adjusted α=0.03, p=0.718, Mann-Whitney U test) or confusion (W=571, adjusted α=0.05, p=0.935, Mann-Whitney U test).

*Focused* also exhibited no significant differences between conditions, with the experimental group's median (Mdn=0.538) not statistically different from the control group (Mdn=0.5; W=537, adjusted α=0.01, p=0.620, Mann-Whitney U test).

## 4. Discussion and summary

In this study, we utilized the GPT-3.5 model to generate personalized hints for students working on programming assignments within an automated assessment platform. Our findings indicated that almost half of the students (46%) highly valued the usefulness of GPT-generated hints, with a median rating of 4 or 5 on a 5-point Likert scale. Given that the assignments and generated hints were provided in Polish, these findings hold promise for the future development and scalability of such systems across various languages, further extending their applicability and impact in diverse educational contexts.

However, 19% of the students found the hints to be not useful, with median ratings of 1 or 2 on the same scale. Thus, there is still room for improvement and for making the hints more useful. Considering the varied ratings of hint usefulness, we anticipate that further optimization of the hint generation process could enhance the efficacy of GPT-based automated feedback, ultimately leading to improved student performance. Future research may be able to further optimize and enhance the hints, to increase the proportion of useful hints and the degree to which each hint is useful. Potential strategies for achieving this include more experimentation with prompt generation (prompt engineering), fine-tuning the GPT model (by providing examples on which the model can improve) or conducting correlation mining to identify the properties associated with less useful hints.

Another finding is that the experimental group (with GPT hints enabled) relied less on the platform's regular feedback for tasks where the GPT feedback was enabled. Despite of significantly lower usage of this kind of feedback, the experimental condition students performed significantly better on the platform in terms of percentage of successful submissions across consecutive attempts. There was not a significant difference in the time needed to successfully solve each assignment with GPT feedback enabled. However, students in the experimental group spent an additional average of 40 seconds after each attempt compared to the control group. This extra time was used by the students to read and understand the hint that was generated before they proceeded to rate it through a survey.

Furthermore, when GPT hints were made unavailable, students in the experimental group needed significantly less time to successfully solve each assignment. Since hints enabled for earlier tasks not only served to rectify the students' misconceptions but also provided supplementary information, it is our conjecture that these hints played a pivotal role in enhancing the students' learning process, apparently further facilitating their understanding. This knowledge acquisition, in turn, may have expedited their task-solving abilities, leading to the observed reduction in time required to complete tasks when the hints were no longer available.

An unexpected outcome of this study is that when students in the experimental condition switched over to more difficult tasks for which GPT hints were not enabled, these students were significantly less likely to successfully solve the tasks in their first several attempts. Our findings indicate that users appeared to rely on the GPT-generated feedback during the study and needed some time to adapt to a lack of this kind of feedback. This may indicate that students were becoming overly dependent on the GPT-generated feedback, but if so, this problem seemed to correct itself fairly rapidly.

Although the GPT hints impacted student performance, the availability of these hints did not appear to have a substantial impact on student affect.

This study presents several limitations that warrant further investigation. First, the study duration spanned only 9 weeks of the 15-week semester term; future research should explore the longer-term effects of GPT-generated hints on student performance and learning outcomes. Second, the experiment focused solely on GPT-generated hints in the Polish language, raising the need for additional experiments in different languages to establish the generalizability of the results. Third, the number of participants represents a limitation to the robustness and external validity of our findings; future work should examine this type of intervention with a larger sample size. Finally, the study utilized a single programming language (C#), while introductory computer science education often incorporates other languages, such as Java and C++. Future research should expand the scope to include a broader range of programming languages to more comprehensively assess the efficacy and applicability of GPT-generated hints.

Addressing the challenge of generating personalized feedback for programming assignments is a demanding task, given the myriad of potential code syntax errors and code complexity across different assignments. The GPT model holds the potential to tackle this issue more effectively and efficiently than human authoring of hints. In summary, the findings of this study indicate that integrating GPT-generated feedback into computer programming education may positively impact student performance, ultimately contributing to enhanced learning outcomes.

Although this experiment employed the default GPT-3.5 model, it is plausible that fine-tuning the model could further improve the quality of generated hints, thereby facilitating more effective programming education on a larger scale.

## Acknowledgements

This paper was written with the assistance of ChatGPT, which was used to improve the writing clarity and grammar of first drafts written by humans. All outputs were reviewed and modified by two human authors prior to submission.